\documentclass[USenglish,twocolumn]{article}

\ifx\directlua\undefined\ifx\XeTeXcharclass\undefined
  \usepackage[utf8]{inputenc}                           %pdftex engine
  \else\RequirePackage[no-math]{fontspec}[2017/03/31]\fi %xetex engine
  \else\RequirePackage[no-math]{fontspec}[2017/03/31]\fi %luatex engine
\usepackage[sort&compress,square,numbers]{natbib}
\usepackage[big,online]{dgruyter}
\usepackage{lineno}
\theoremstyle{dgthm}

\theoremstyle{dgdef}

\begin{document}

\articletype{Letter}

% \author[1]{First author}
% \author[2]{Second author}
% \author*[3]{Third author} 
% \affil[1]{University, City, Country, e-mail; ORCID}
% \affil[2]{University, City, Country, e-mail; ORCID}
% \affil[3]{University, City, Country, e-mail; ORCID}

\author*[1]{Marko Perestjuk}
\author[2]{Rémi Armand}
\author[2]{Miguel Gerardo Sandoval Campos}
\author[2]{Lamine Ferhat}
\author[4]{Vincent Reboud}
\author[4]{Nicolas Bresson}
\author[4]{Jean-Michel Hartmann}
\author[4]{Vincent Mathieu}
\author[3]{Guanghui Ren}
\author[5]{Andreas Boes}
\author[3]{Arnan Mitchell}
\author[2]{Christelle Monat}
\author[2]{Christian Grillet}

\affil[1]{Ecole Centrale de Lyon, INSA Lyon, CNRS, Universite Claude Bernard Lyon 1, CPE Lyon, INL, UMR5270, 69130 Ecully, France \& School of Engineering, RMIT University, Melbourne, VIC 3001, Australia; E-mail: marko.perestjuk@ec-lyon.fr}
\affil[2]{Ecole Centrale de Lyon, INSA Lyon, CNRS, Universite Claude Bernard Lyon 1, CPE Lyon, INL, UMR5270, 69130 Ecully, France}
\affil[3]{School of Engineering, RMIT University, Melbourne, VIC 3001, Australia}
\affil[4]{Université Grenoble Alpes, CEA-Leti, 38054 Grenoble, France}
\affil[5]{Institute for Photonics and Advanced Sensing \& School of Electrical and Mechanical Engineering, The University of Adelaide, Adelaide, SA 5005, Australia}

\title{One Million Quality Factor Integrated Ring Resonators in the Mid-Infrared}
\runningauthor{M. Perestjuk et al.}
\runningtitle{One Million Quality Factor Integrated Ring Resonators in the Mid-Infrared}
\abstract{We report ring resonators on a silicon germanium on silicon platform operating in the mid-infrared wavelength range around 3.5 - 4.6 µm with quality factors reaching up to one million. Advances in fabrication technology enable us to demonstrate such high Q-factors, which put silicon germanium at the forefront of mid-infrared integrated photonic platforms. The achievement of high Q is attested by the observation of degeneracy lifting between clockwise (CW) and counter-clockwise (CCW) resonances, as well as optical bistability due to an efficient power buildup in the rings.}
%\journalname{Nanophotonics}
%\dedication{our professor, Dr. Firstname Lastname, whose critique of the underlying study identified potential bias in the analysis and strengthened our argument.}
\journalyear{2025}
\journalvolume{vol}

\maketitle

\vspace*{-6pt}

\section{Introduction} 
\vspace*{-3pt}

Mid-infrared (MIR) photonics, encompassing the wavelength range of around 3 to 15 µm, has generated significant attention due to its relevance in numerous fields. It is characterized by strong fundamental absorption lines of many molecules, making it highly suitable for applications in chemical sensing, environmental monitoring, medical diagnostics and defense \cite{Soref2010}. Integrated photonics is pivotal in advancing MIR technologies. The development of photonic integrated circuits (PICs) in the MIR facilitates the fabrication of compact, efficient, and scalable devices. Such advancements are critical for the deployment of MIR technologies in real-world settings, where the demand for high-precision, portable, and cost-effective solutions is ever-increasing. While the existence of strong absorption lines is the main merit of the MIR, it also makes integrated MIR photonics technologically challenging. Indeed, several materials used in common photonic platforms are also absorbing in the MIR and are therefore not attractive. A plethora of different platforms have been suggested and investigated for the MIR \cite{Lin2018}. One very promising class are group-IV photonic materials, such as silicon germanium alloys (SiGe) \cite{Soref2010}. They have attracted  attention in recent years due to their large transparency window, ability to be used for low loss waveguides, strong nonlinearity and CMOS compatibility \cite{Soref2010}. A fundamental and important building block of PICs is a compact resonator, which can efficiently confine light, i.e. with a high quality factor Q. Microrings are vital in PICs for the design of optical filters, modulators and optical signal processing. Their biggest merit in the MIR is for a use as sensors \cite{Bogaerts2011} and for frequency comb generation \cite{Chang2022}. For both of these key applications, a high Q is crucial as well as a compact mode volume V. When used as an evanescent field sensor, the spectral signature of ring resonators can be changed by an analyte and a higher Q translates into higher sensitivity. For the generation of Kerr frequency combs a high Q$^2$/V is also required as the threshold power scales with the inverse square of Q and is proportional to V. While group-IV integrated photonics has enabled a lot of progress in integrated MIR photonics, such as the demonstration of octave spanning supercontinuum generation \cite{Sinobad2018}, it is lagging behind in the domain of high-Q cavities. It was only very recently that our group demonstrated a high-Q ring resonator on SiGe/Si with a Q of 236,000 \cite{Armand2023} and a Q of 154,000 on Ge/Si \cite{Armand_CLEO2023} around 4 µm wavelength. The results on high-Q resonators on SiGe- and Ge-based  platforms by other groups show the community's great interest in this quest \cite{Koompai2023},\cite{Lim2023}.

In this letter, we demonstrate a significant improvement of the Q-factor of SiGe/Si ring resonators towards one million around 4 µm wavelength, which is a record for group-IV photonics. This was achieved thanks to an optimized fabrication process. Furthermore, we use a simple characterization method that allows a precise Q-factor measurement without the need for any complex instrumentation on the detector side as well as the ability to probe a wide range of resonances with less constraints on the coupling regime. Thanks to this, we perform an analysis of the statistical distribution of Q-factors in a relatively broad part (3.5~-~4.6 µm) of the MIR band. With a Q factor of one million, this puts SiGe/Si ahead of other integrated MIR photonic platforms in the quest for high-Q. The high Q achieved in these SiGe/Si rings allows us to observe clockwise and counter-clockwise resonance splitting \cite{Kippenberg2002},\cite{Bogaerts2011},\cite{Zhu2009} and optical bistability \cite{Almeida2004},\cite{Quan2011}, which typically require high Q-factors.

\vspace*{-13pt}

\section{Design and Fabrication}
\vspace*{-4pt}

\begin{figure} 
\centering
\includegraphics[width=1.0\linewidth]{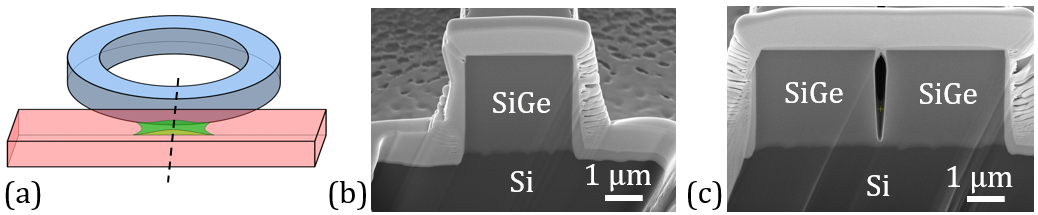}
\caption{\label{fig:FIB} Fabrication improvement of ring resonators. (a) Schematic showing that for small gaps and high aspect ratio, an undesirable residue of the waveguide material can be left in the gap. (b) Scanning-electron microscope (SEM) image of SiGe-on-Si waveguide. (c) SEM image obtained after fabrication improvement, showing a full etching of a 250 nm gap.}
\end{figure}

We designed resonators for a platform of air-cladded Si$_{0.6}$Ge$_{0.4}$ core waveguides on a Si substrate. Details on the platform and the resonator design can be found in Refs. \cite{Sinobad2018},\cite{Armand2023}. The composition was chosen as a compromise between higher index contrast and transparency (for higher Ge content) and lower lattice mismatch with the Si substrate (for lower Ge content). The refractive index of SiGe at 4 µm wavelength is n = 3.57 and for the Si substrate n = 3.42. The waveguide cross-section dimensions were chosen based on mode guiding considerations (single-mode/multi-mode) as well as dispersion. We targeted anomalous dispersion to make the resonators more attractive for nonlinear applications. Based on these considerations, the waveguide height was set to h = 3.3 µm. At this height and a wavelength of $\lambda$ = 4 µm for example, the fundamental TE mode is cut off at a waveguide width lower than around 2.2 µm, the waveguide is single-mode at 2.2 - 4.5 µm width and multi-mode beyond 4.5 µm width. Propagation losses as low as 0.15 dB/cm have been demonstrated on this platform \cite{Torre2022} for straight waveguides. The ring radius was chosen to be R = 250~µm where bending losses become negligible \cite{Armand2023} and the resonator is relatively compact (i.e. optimizing Q/V). Gaps between rings and straight bus waveguides were at a minimum of g = 250 nm (the smallest gap manufacturable with the used process) and increasing in steps of 250~nm to have some variation of the coupling efficiency. 

The fabrication was performed on a 200 mm CMOS pilot line at CEA-Leti. SiGe was grown epitaxially on Si and then patterned by deep ultraviolet lithography followed by deep reactive ion etching  \cite{Sinobad2018}. The fabrication of the small gap is technologically challenging due to the high aspect ratio, e.g. 13.2 for h~= 3.3 µm and g = 0.25 µm. The resulting gap was imaged by making a cross-section cut at the gap using a focused ion beam (see Fig. \ref{fig:FIB}). In previous fabrication runs, the 250 nm gap was not fully etched as illustrated in Fig. \ref{fig:FIB}a, which can be a potential source of additional loss. For the batch of resonators presented here, the fabrication and specifically the etching process was consequently optimized, especially the resist selectivity and the etching time control. The resulting 250 nm gap is shown in Fig. \ref{fig:FIB}c with an improved result. 

\vspace*{-13pt}

\section{MIR Characterization}
\vspace*{-4pt}

\begin{figure}
\includegraphics[width=1.0\linewidth]{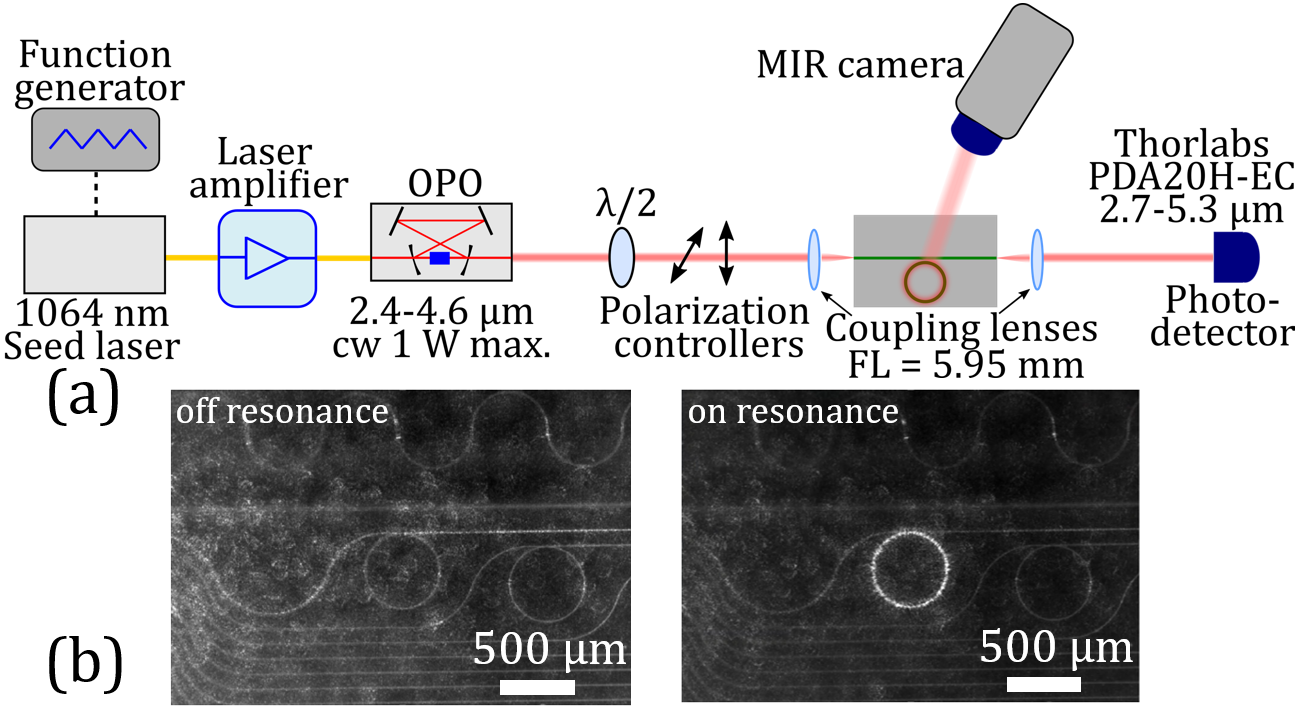}
\caption{\label{fig:Setup} (a) MIR measurement setup (FL = focal length). (b) MIR camera images showing the light scattered by a ring resonator when excited off (left) and on resonance (right).}
\end{figure}

We used the measurement setup schematically shown in Fig.~\ref{fig:Setup}a to characterize the resonators. The source was a tunable continuous wave optical parametric oscillator (OPO) from Excelitas (iFLEX-Agile cw-OPO, idler $\sim $2.4 - 4.5 µm, $\sim$10 MHz linewidth), which is pumped by an amplified near infrared seed laser. Modulating the seed laser gives fine tuning of the OPO output wavelength and hence allows us to scan through the ring resonance continuously. A function generator fed into the piezo controller of the seed laser enables a continuous tuning and modulation up to kHz speed over a 13 GHz range ($\sim$700~pm at $\lambda$ = 4 µm). A wave plate and polarization controllers were used for setting the input to TE polarization and for controlling the input power. For the characterization of the Q factors, a relatively low off-chip input power of 10 mW was used. Beyond the standard transmission measurements \cite{Armand2023}, we developed a setup where we also image the light scattered off the cavity in the vertical direction using a MIR camera in top-view (FLIR, A6750 MWIR) with a MIR objective lens (0.2 numerical aperture). We can reasonably assume that the detected scattered light is directly proportional to the intracavity power, since scattering is mediated by the sidewall roughness of the ring, i.e. it is relatively isotropic and wavelength-independent. The signal recorded by the camera holds a linear relationship to the detected power. With this approach we have several advantages. First and foremost, this allows us to measure resonances for a larger range of coupling gaps, providing that the intensity build-up in the resonator is high enough (total Q factor). Even when no resonance is visible in transmission, we are able to observe the power build up in the resonator in top-view. Furthermore, the Fabry-Perot fringes formed by the chip end facets \cite{Armand2023}, which overlap and can obscure the ring resonances on transmission spectra, are less of an issue on the scattered light recording provided by the camera. Fig.~\ref{fig:Setup}b shows an example of a ring resonator off and on resonance when changing the laser wavelength from 4106.0 nm to 4106.1 nm. Off resonance, light scattered by the input bus waveguide can be observed as well as an overall noise background due to light scattering induced by the coupling to the chip input facet. On resonance, the additional light scattered by the ring is clearly visible with a high signal to noise ratio, attesting the occurrence of coupling for selected wavelengths.  

\begin{figure*} 
\centering
\includegraphics[width=1.0\textwidth]{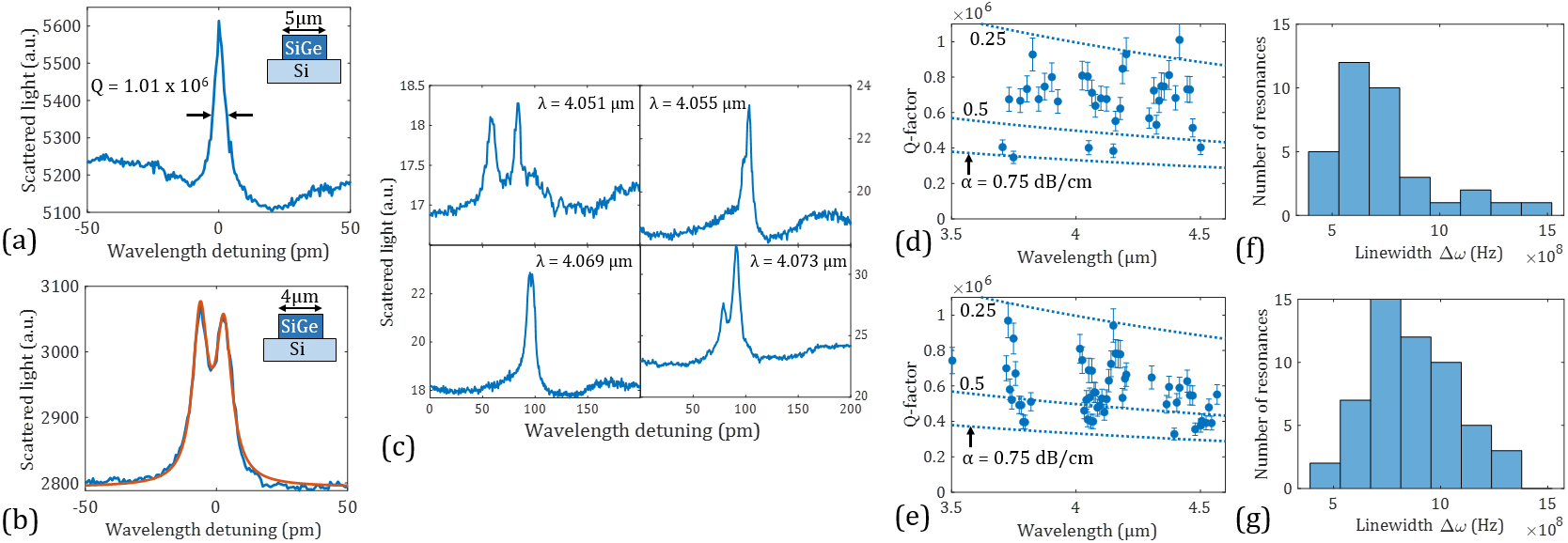}
\caption{\label{fig:Histograms} Q-factor measurements for waveguide widths of 5 µm (a,d,f) and 4 µm (b,c,e,g). (a) and (b) show examples of measured resonances. In (a) a record Q-factor of one million is measured at a wavelength of 4.42 µm, while in (b) a resonance splitting can be observed for a wavelength of 4.57 µm. The brown line is a fit according to the model from Ref. \cite{Zhu2009}. (c) Different resonance splittings are observed for different resonances on the same ring (4 µm width). (d) and (e) show the measured Q-factor for different resonances. The dashed lines indicate the corresponding propagation losses assuming that those are the sole loss contribution. (f) and (g) show the histograms of resonance linewidths.}
\end{figure*}

With this setup we measure the total Q of resonances. As these can be detected for strongly under-coupled regimes, we effectively have a direct way to probe the intrinsic Q-factor of the ring. This becomes apparent from the following relationship
\begin{equation}
\frac{1}{Q} = \frac{1}{Q_{in}} + \frac{1}{Q_c},~Q_{in} = \frac{10}{ln(10)}\frac{2 \pi n_g}{\alpha \lambda},~Q_c =\frac{2 \pi n_g L}{\kappa^2 \lambda}
\label{eq:Q}
\end{equation}
where $Q$ is the total Q-factor, $Q_{in}$ the intrinsic Q-factor, $Q_c$ the coupling Q-factor, $\alpha$ the propagation loss in dB/m, $n_g$ the group index, $L$ the resonator length and $\kappa^2$ the power coupling coefficient. If the resonator is strongly under-coupled, then $\kappa$ becomes very small and consequently $Q_c$ very large. Then, because of $Q_c >> Q_{in}$, we get $Q \approx Q_{in}$, i.e. while the measurement is always giving $Q$, this is equivalent here to $Q_{in}$. Accordingly, we focus our measurements on a gap of 500 nm, where rings are strongly under-coupled and no resonances are visible in transmission but still clearly visible on the camera. Under-coupling is also expected for this gap from performed simulations (Lumerical FDE and FDTD). We probe rings with waveguide widths of 4 µm and 5 µm (ring and bus), so we have a comparison between single-mode and multi-mode waveguides, while probing the trade-off between distributed loss (see Ref. \cite{Torre2022}) and coupling to the bus waveguide. Fig. \ref{fig:Histograms} shows the measurement results for the two different widths. In Fig. \ref{fig:Histograms}a an example of a high Q resonance at $\lambda$ = 4.42 µm is shown. The extracted Q from the full width at half maximum ($Q = \lambda / \Delta \lambda$) is $(1.01 \pm 0.10) \times 10^6$.

In the spectrum in Fig. \ref{fig:Histograms}b a typical example of resonance splitting can be observed, which we attribute to CW and CCW degeneracy lifting \cite{Kippenberg2002}. The underlying CW and CCW mode coupling could be caused by distributed scattering at the sidewalls or scattering at a defect, particle or the coupler itself \cite{Bogaerts2011},\cite{Zhu2009}. This mode splitting generally attests a high Q because the coupled resonances get otherwise spectrally smeared out. We note that other phenomena such as a TE/TM coupling would lead to further spectral separation, inconsistent with our measurements. In Fig. \ref{fig:Histograms}c examples of the splitting for different resonances are shown. Doublet splitting can express itself with different strengths, sometimes giving a clear splitting, just a shoulder or no splitting at all. We use the model from Ref. \cite{Zhu2009} to fit Fig.~\ref{fig:Histograms}b data. The resulting key parameters are a $Q_{in}$ of 674,000 and a doublet splitting of 0.13 GHz. Resonance splitting in high-Q systems has also been harnessed in sensing applications, where the splitting is induced by an analyte \cite{Vollmer2012}.

A systematic study of the resonance Q factors was performed on a large number of resonances over a wide wavelength range (3.5 - 4.6 µm). Fig. \ref{fig:Histograms}d and e show that the Q-factor varies over a factor of more than 2 for the same ring for both waveguide widths. This is statistically expected and similar or even higher spreads are typically observed (e.g. see Refs. \cite{Ye2019},\cite{Kordts2016}). Q also tends to decrease for longer $\lambda$, which can be attributed to the $\lambda$-dependence of $Q_{in}$. To illustrate this, the dashed lines in Fig.~\ref{fig:Histograms}d-e show the expected $Q_{in}$ for constant propagation loss. If we assume that this is the dominant loss contribution, the majority of measured Q values correspond to propagation losses between 0.2 and 0.5 dB/cm, in line with the loss measurements of straight waveguides \cite{Torre2022}. As only resonances with the highest Q reach the typical propagation loss around 0.2 dB/cm measured for this platform, this means that the "best" resonances are dominated by propagation loss, while lower Q resonances are most likely affected by other contributions. Histograms of resonance linewidths in Fig. \ref{fig:Histograms}f-g resemble typically observed Burr distributions \cite{Kordts2016}. The linewidth is plotted as it factors out the expected decrease of the Q factor for increasing $\lambda$, assuming constant loss. It can be seen that a 5 µm waveguide width yields slightly higher Q-factors compared to 4 µm. This means that the onset of multi-mode guiding is not deteriorating the Q. Instead, the lower propagation loss due to less interaction with the sidewalls in the wider waveguide effectively translates into a higher Q. The resonance splitting was observed to be generally stronger for the 4 µm wide waveguide, suggesting that  splitting is induced by distributed scattering \cite{Bogaerts2011}.

\begin{figure}
\centering
\includegraphics[width=1.0\linewidth]{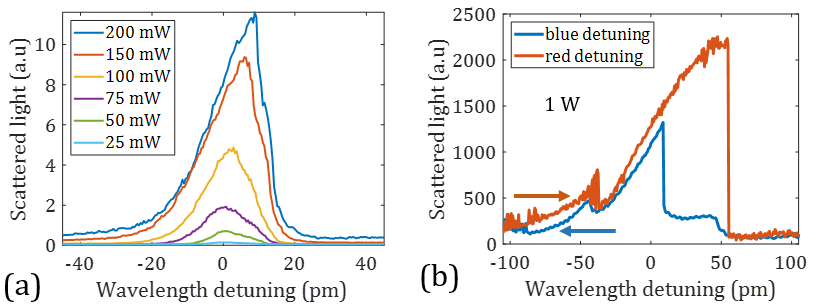}
\caption{\label{fig:Bistability} Measured resonance for increasing (off-chip) input pump power (Q = 280,000). As the power is increased in (a) a progressive tilt is observed. Optical bistability is reached in (b) for a further power increase, also indicated by the difference between $\lambda$-tuning direction.}
\end{figure}

\begin{table*}
\centering
\caption{Comparison of the intrinsic Q-factors achieved for different integrated photonic platforms in the MIR.}
\begin{tabular}{ p{1.4cm}p{3cm}p{2cm}p{2cm}p{2.5cm}p{3cm} }
Ref. & Platform & Record (intrinsic)  Q-factor & Wavelength (µm) & Max. wavelength of platform (µm) & Mode volume  V/2$\pi$ = R x A (µm x µm$^2$)\\ \midrule
\cite{Luke2015} & SiN-on-SiO$_2$ & $10^6$ & 2.6 & 3.5 & 230 x 2.6 \\
\cite{Miller2017} & Si-on-insulator & $10^6$ & 3.5 - 3.8 & 3.8 & 150 x 9.2 \\
\cite{Shankar2013} & Si-on-sapphire & 278,000 & 4.3 - 4.6 & 5.5 & 60 x 1.2\\
\cite{Miller2016} & Suspended Si & 83,000 & 3.8 & 8.5 & 150 x 0.4 \\
\cite{Lin2014} & Chalcogenides (GeSbS) & 550,000 & 5.2 & 11 & 500 x 2.7 \\
\cite{Karnik2023} & InGaAs/InP & 174,000 & 5.2 & 14 & 500 x 6.2 \\
\cite{Haas2019} & GaAs/AlGaAs & 1,900 & 5.5 - 5.9 & 15 & 248 x 36 \\
\cite{Lim2023} & Ge-on-Y$_2$O$_3$ & 176,000 & 4.2 & 13 & 63.5 x 0.6 \\
\cite{Koompai2023} & Graded SiGe-on-Si &  113,000 & 7.5 - 9.0 & 15 & 250 x 16 \\
\cite{Armand2023} & SiGe-on-Si & 236,000 & 4.2 & 8.5 & 250 x 13.2 \\
This work & SiGe-on-Si & $10^6$ & 3.5 - 4.6 & 8.5 & 250 x 13.2 \\
\end{tabular}
\label{tab:Q-comparison}
\end{table*}

Finally, we investigate the behavior of resonators under strong power loading. For this, we increase the power and the coupling strength by investigating rings with a smaller gap of 250 nm and a smaller waveguide width of 3.25 µm, operating around $\lambda$ = 4.1 µm. The Q-factor of the chosen resonance lowered by loading is 280,000. In Fig. \ref{fig:Bistability} we see a clear signature of optical bistability for increasing power. Bistability can be seen when the power buildup in a cavity is large enough to significantly change the refractive index of the cavity material \cite{Almeida2004}. This leads to a feedback loop and a progressive “tilt” of the resonance, finally resulting in a two state system. For a given pump wavelength there are two resonance states, so the cavity can act as a bistable switch. Fig. \ref{fig:Bistability}a shows such a slight increasing tilt of the resonance for increasing power. In Fig. \ref{fig:Bistability}b bistability is clearly observed under higher power, where distinct spectral signatures are recorded upon red- or blue-tuning the laser wavelength across the resonance spectrum. Bistability may be attributed to the thermo-optic effect or the Kerr nonlinearity. A method to determine the nature of the bistability is to use a fast modulation of the input power. Unlike the instantaneous Kerr effect, the thermo-optic effect is on µs time scales, so a MHz modulation would remove the latter. In the MIR however, fast modulators are not commonly available. Regardless of the nature of the bistability, our results show that SiGe/Si rings can efficiently store a high circulating power. Besides being a precursor for the use of other nonlinearities, bistability itself has many applications. It can for instance be used in all optical switching \cite{Almeida2004}, and sensing applications, as the steep edge of the resonance results in a strong sensitivity enhancement \cite{Quan2011}.

\vspace*{-13pt}

\section{Discussion and Conclusions} 
\vspace*{-4pt}

The demonstrated Q of $10^6$ is a record value for SiGe- and Ge-based integrated photonics in the probed wavelength range of 3.5 - 4.6 µm. It is among the highest Q-factors in the MIR across all integrated platforms. The measurement uncertainty stems from the precision to which the $\lambda$-modulation range can be determined as well as statistical uncertainties between measurements based on wavelength and temperature fluctuations. Tab. \ref{tab:Q-comparison} shows a comparison of the highest Q-factors for different integrated photonic platforms in the MIR. Similar Q-factors of $10^6$ were scarcely achieved and only at the very start of the MIR range ($\lambda$ < 3.8 µm). More importantly, this was achieved with silica-based platforms which are intrinsically limited in wavelength as silica absorbs beyond 3.5 µm. Hence, SiGe/Si is now the first integrated platform that reaches Q's of one million and which covers most of the MIR range. Previously, with our SiGe/Si platform we have demonstrated operation up to at least 8.5 µm \cite{Sinobad2018}, where transparency is generally limited by substrate absorption. Considering that the exact transparency range of SiGe alloys depends on the Ge content and that interactions with the Si substrate could be reduced in highly confining waveguides, extension up to 15 µm could be envisaged \cite{Sinobad2018},\cite{Koompai2023}.

In conclusion, we have demonstrated a record Q-factor for integrated photonics in the MIR of one million in a SiGe/Si platform. Application fields include sensing and nonlinear integrated photonics, in particular Kerr comb generation. The high Q also manifested itself in the appearance of resonance splitting and optical bistability. Bistability can be useful to increase sensitivity in sensing applications.

\begin{funding}
We acknowledge the support of the Horizon 2020 research and innovation program under the Marie Sklodowska-Curie Actions (Grant No. ECLAUSion, 801512), the Agence Nationale de la Recherche (ANR) (Grant No. MIRSiCOMB, ANR-17-CE24-0028; MIRthFUL, ANR-21-CE24-0005), the Australian Research Council Centre of Excellence in Optical Microcombs for Breakthrough Science (CE230100006) and the International Associated Laboratory in Photonics between France and Australia (LIA ALPhFA). This work benefits from a France 2030 government grant managed by the French National Research Agency (PEPR OFCOC, ANR-22-PEEL-0005).
\end{funding}

\begin{authorcontributions}
All authors have accepted responsibility for the entire content of this manuscript and consented to its submission to the journal, reviewed all the results and approved the final version of the manuscript. MP, RA performed the design, experimental characterization and conception of the setup. MGSC and LF performed the experimental characterization. VR, NB, JMH, VM conducted the fabrication of the samples. GR, AB, AM, CM and CG conceived the idea and supervised the project. MP prepared the manuscript with contributions from all co-authors.
\end{authorcontributions}

\begin{conflictofinterest}
Authors state no conflict of interest.
\end{conflictofinterest}

\begin{informedconsent}
Informed consent was obtained from all individuals included in this study.
\end{informedconsent}

\begin{dataavailabilitystatement}
The datasets generated and analyzed during the current study are available from the corresponding author on reasonable request.
\end{dataavailabilitystatement}

\vspace*{-22pt}

\bibliographystyle{ieeetr}
\bibliography{bibliography}

\clearpage

\end{document}